\RequirePackage{lineno}
\documentclass{camera}
\usepackage{graphicx}  % uncomment this if your want to insert figures
\usepackage{amssymb}
\input epsf
\begin{document}

%%%%%%%%%%%%%%%%%%%%%%%%%%%%%%%%%%%%%%%%%%%%%%%%%%%%%%%%
% The title, only the first letter capitalized; if you want to split it in
% two or more lines, put a \\ macro at each line break
% example: 
%   \title{Title: first line\\ second line}
%
\title{Search for neutrino point sources with the IceCube Neutrino Observatory.}

%%%%%%%%%%%%%%%%%%%%%%%%%%%%%%%%%%%%%%%%%%%%%%%%%%%%%%%%
% The author(s), separated by commas; do not put a
% comma before the last author, use instead the \and
% macro which produces a normal ``and'' in the
% caps/small caps context
%
\author{Juan A. Aguilar$^{*}$ for the IceCube Collaboration$^{\dagger}$.}

%%%%%%%%%%%%%%%%%%%%%%%%%%%%%%%%%%%%%%%%%%%%%%%%%%%%%%%%
%
\organization{$^{*}$Dept. of Physics, University of Wisconsin, Madison, WI 53706, USA
\\ $^{\dagger}$http://icecube.wisc.edu}
\maketitle

\begin{abstract}

The IceCube Neutrino Observatory is a kilometer-scale detector currently under construction at the South Pole. The full detector will comprise 5,160 photomultipliers (PMTs) deployed on 86 strings from 1.45-2.45 km deep within the ice. As of the austral summer of 2009-10, 73 out of the total number strings have been deployed, and the detector is reaching its final construction phase. A dense sub-array of 6 strings in the center of the detector (DeepCore) has been already installed for enhancing the sensitivity to low energy neutrinos. The IceCube detection principle is based on the measurement of the Cherenkov light induced by ultra-relativistic muons and showers produced by neutrino interactions in the target matter of the detector. The main scientific goal of the IceCube experiment is the detection of astrophysical neutrinos that will help to understand and settle the unresolved questions about the origin and nature of cosmic rays. In this contribution we will present the latest results of the experiment concerning the search for neutrino point sources using the experimental data taken during 2008-09 where the detector was operated with a 40-string configuration.  The results of the analysis for steady individual neutrino sources as well as the stacking analysis from different catalogs will be presented.

\end{abstract}

%%%%%%%%%%%%%%%%%%%%%%%%%%%%%%%%%%%%%%%%%%%%%%%%%%%%%%%%
% Write the text starting from here and using the usual
% LaTeX commands.
%

\section{Introduction}

The IceCube Neutrino Observatory is a neutrino telescope designed to detect high energy astrophysical neutrinos with energies $\gtrsim~100$~GeV. Such an observation could reveal the origins of cosmic rays (CR) and the possible connection to shock  acceleration in Supernova Remnants (SNR), Active Galactic Nuclei (AGN) or Gamma Ray Bursts (GRB). An advantage of neutrino astronomy over other experimental observations such as gamma ray astronomy is the possibility to offer insight about the most energetic and dense parts of the Universe since neutrinos are weakly interactive particles that can travel through matter without interacting. Nonetheless, this advantage also makes the neutrino detection a technological challenge since a large amount of volume is required to observe neutrino interactions. The IceCube Neutrino Observatory uses the Antarctic ice as a detection volume where a neutrino interaction will produce a muon (or other charged lepton) as a result of the charged current interaction. Deep Antarctic ice is also transparent which allows the induced Cherenkov light by the muon to propagate. In order to detect this Cherenkov light, IceCube will instrument a cubic kilometer of clear Antarctic ice sheet underneath the geographic South Pole with an array of 5,160 Digital Optical Modules (DOMs) deployed on 86 strings from 1.45-2.45 km deep. The DOMs are spherical, pressure resistant glass housing, each containing a 25 cm diameter Hamamatsu photomultiplier tube and the electronics for waveform digitization. 
The IceCube configuration also includes a denser array with 6 strings equipped with higher quantum efficiency DOMs, the so called DeepCore detector, in order to increase the sensitivity to low energy neutrinos ($\lesssim$~100~GeV). A surface array for observing extensive air showers of cosmic rays is also being installed (IceTop). IceCube construction started with a first string installed in the 2005-6 season (Achterberg et al. 2006) and will be completed in the austral summer of 2010-11.

In this paper we describe the point source analysis of the data corresponding to the year 2008-9 when IceCube configuration consisted in 40 deployed strings. The IceCube 40-string configuration performance as well as the event selection criteria followed in the analysis are described in section~\ref{sec:performance}. Section~\ref{sec:method} explains the methodology used for the search of point neutrino sources and stacking analysis. The search strategies applied to the IceCube data and their results are discussed in section~\ref{sec:searches}. Conclusions are shown in section~\ref{sec:conclusions}.

\section{Event selection and detector performance}
\label{sec:performance}

As mentioned before, forty strings of IceCube were operational from April 2008 to May 2009 with $\sim 90\%$ duty cycle after a run selection based on the stability of the detector. The number of trigger events of the IceCube detector for the 40-string configuration livetime was of the order of $\sim 3 \times 10^{10}$ events. Among these events, 99.99999$\%$ are muons produced by the impact of primary cosmic rays with the atmospheric nuclei. This number of triggered events can be reduced to $\sim 8 \times 10^{8}$ events using an online filtering system based on the quality of the track reconstructions and energy estimators. The total livetime of the 40-string configuration was 375.5 days, leading to final atmospheric neutrino rate of 40 atmospheric neutrinos per day, which comprises the irreducible background of a point source analysis. The detection principle of neutrino telescopes normally uses the Earth as a shield for up-going atmospheric muon tracks. By selecting only reconstructed up-going tracks we can reject part of these atmospheric muons. However due to the huge amount of down-going atmospheric muons reaching the detector, even a small fraction of mis-reconstructed events can contaminate the up-going sample. For that reason tight cuts in the quality of the track reconstruction and track-like parameters such as the reduced likelihood of the track fit and the directional width of the likelihood space around
the best track fit are applied. Also a cut in the likelihood ratio between the best up-going and down-going track solution is applied together with a requirement that the event's set of hits can be split into two parts which both reconstruct as nearly up-going. These cuts in the up-going region (northern sky from IceCube location) define a final data sample with sensitivity to point sources optimized for sources of neutrinos in the TeV-PeV energy range.

With the 40-string configuration we also extended the point source analysis to directions above the horizon in the down-going region of the sky (southern sky from the IceCube location). This extension of the point source analysis is possible by rejecting five orders of magnitude of atmospheric muon background events by applying cuts in the energy estimator in order to discriminate the large amount of these atmospheric muons from a possible neutrino signal with a harder spectrum. After track-quality selections, similar but tighter than for the up-going sample, a cut based on an energy estimator is made until a fixed number of events per steradians is achieved. Because only the highest energy events pass the selection, sensitivity in part of the sky is primarily to neutrino sources at PeV energies and above.

% For Figures insertion uncomment the next section
\begin{figure}[ht!]  %%% FIGURE 1 %%%
\begin{center}
\epsfysize=6cm \epsfbox{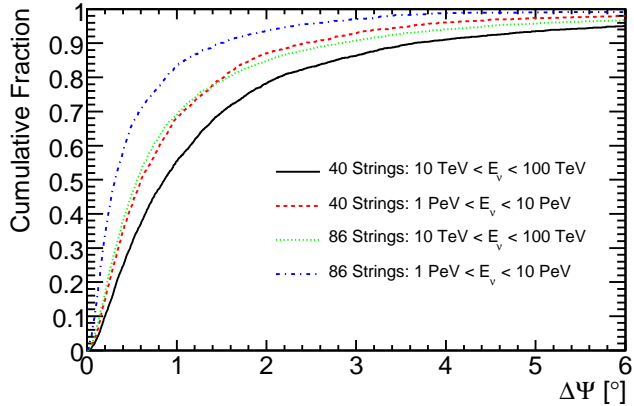} 
\vspace{0.3cm}
\caption[h]{\label{fig:psf} Cumulative point spread function. The horizontal axis represents the angle between the true neutrino direction and the reconstructed muons. The vertical axis is the fraction of events with an angular distance less or equal to the corresponding value on the horizontal axis.}
\end{center}
\end{figure}

% For Figures insertion uncomment the next section
\begin{figure}[ht!]  %%% FIGURE 2 %%%
\begin{center}
\epsfysize=6cm \epsfbox{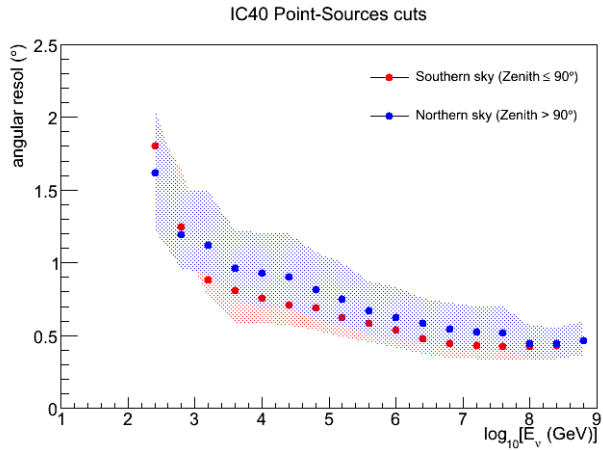} 
\vspace{0.3cm}
\caption[h]{\label{fig:resol} Angular resolution, defined as the median of the point spread function, for selected events in the up-going region (blue) and down-going region (red). The dashed area indicates a 10\% area.}
\end{center}
\end{figure}

Both the angular resolution and effective area of the detector depend on the event selection. Since the two samples (up-going and down-going events) are defined by different quality cuts, the detector performance is different for both parts of the sky. Figure~\ref{fig:psf} shows the neutrino point spread function of the IceCube 40-string configuration for up-going events compared to the past 22-string configuration and the final IceCube detector. The angular resolution, defined as the median of the point spread function, is shown in figure~\ref{fig:resol} for both the up-going and down-going region for the final IceCube 40-string sample. After the event selection described in the previous section, the final data sample consists of 36900 events where approximately one third are up-going events and two thirds down-going events.

% For Figures insertion uncomment the next section
\begin{figure}[ht!]  %%% FIGURE 3 %%%
\begin{center}
\epsfysize=6cm \epsfbox{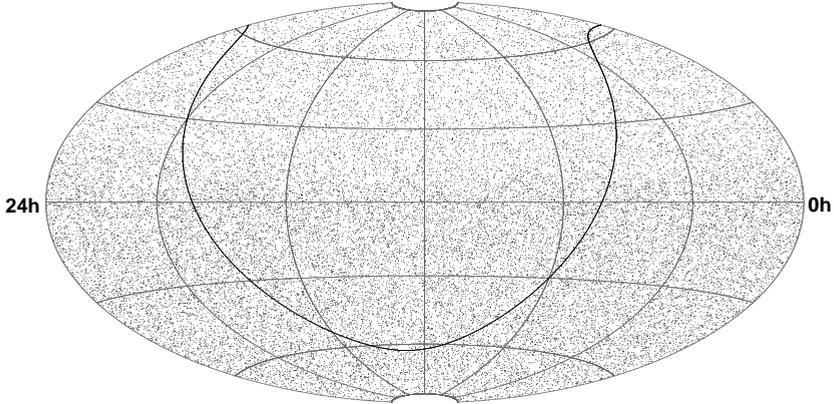} 
\vspace{0.3cm}
\caption[h]{\label{fig:eventmap} Equatorial sky map of 36900 events in IceCube 40-strings after the analysis cuts.}
\end{center}
\end{figure}

\section{Methodology}
\label{sec:method}

The final sample of a point source analysis consists of a set of arrival directions and energy estimators. Figure~\ref{fig:eventmap} shows how the final arrival directions are distributed on the whole sky. The aim of a point source analysis is to identify an anisotropy on these arrival directions by looking for an accumulation of events in a given direction which will indicate the presence of an astrophysical point source emitting neutrinos. In order to perform such an analysis in IceCube we use an unbinned likelihood method (Braun et al. 2008). This method has substantial improvements with respect to the standard binned analysis. The method accounts for individual reconstructed events uncertainties and energy estimators and compares the random distributions of atmospheric neutrino background for the northern sky and atmospheric muon background for the southern sky to that produced by a signal coming from a point neutrino source. The method models the data as a two component mixture of signal and background. A maximum likelihood fit is applied to determine the relative contribution of each component.
Given $N$ events in the final sample the density distribution of the $i^{th}$ event is given by

\begin{equation}
\label{eq:pdf}
\frac{n_{s}}{N}\mathcal{S}_{i} + \left(1- \frac{n_{s}}{N}\right)\mathcal{B}_{i},
\end{equation}

\noindent where $\mathcal{S}_{i}$ is the probability density distribution for the signal hypothesis and $\mathcal{B}_{i}$  for background. The parameter $n_{s}$ is the number of signal events and is one of the free parameters of the likelihood maximization together with the spectral index, $\gamma$, of the signal spectrum distribution.

For each tested direction in the sky, the best fit according to the likelihood maximization is found yielding the best estimate of $\hat{n}_{s}$ and $\hat{\gamma}$. The logarithm of the likelihood ratio between the best-fit hypothesis, $\mathcal{L}(\hat{n}_{s},\hat{\gamma})$, and the null hypothesis of only background, $\mathcal{L}(n_{s} = 0)$ is the test statistic of the method. The significance of the result is evaluated by comparing the test statistic with a distribution obtained by performing the same analysis over a sample of only-background data sets. These only-background datasets are obtained by scrambling real data in right ascension. The Earth rotation guarantees the uniform exposure in right ascension and uniform background per declination band, hence we can randomize the events in right ascension while keeping all other event properties fixed. Nearly vertical events are excluded from the randomization since polar regions in the sky cannot be scrambled in right ascension.

In all-sky analyses the whole sky is scanned in steps of $0.1^{\circ} \times 0.1^{\circ}$ forming  a grid with bin size much finer than the angular resolution. The hottest spot in the grid is considered to be the point with the highest significance. The final significance is determined by the fraction of scrambled data sets containing at least one grid point with a log likelihood ratio higher than the one observed in the data. This fraction is the {\it post-trial} p-value for the all-sky search.

A significant penalty is paid in the all-sky search due to the number of effective trials due to scanning the entire sky. Therefore an additional search usually performed in point source analysis is to select a number of locations {\it a priori} based on candidate sources. The likelihood is maximized on each of these locations instead of over the entire sky reducing the number of trials. The most significant result from this candidate source list is also compared to the distribution of randomized scrambled data yielding the final {\it post-trial} p-value.

A third method of stacking analysis is also presented in this paper where instead of considering individual sources the cumulative signal from a collection of similar sources is considered. The stacking analysis is a well-known technique in astronomy that has been already applied to neutrino astronomy (Abbasi et al. 2009). Stacking multiple sources is an effective way to enhance the discovery potential of similar sources when the source fluxes are too faint to yield a discovery individually. By considering the cumulative signal for all sources emitting together, we can increase our discovery potential and set upper limits on interesting catalogs of neutrino source candidates as a whole. The stacking analysis requires a small modification to the unbinned likelihood analysis. The signal density distribution, $\mathcal{S}_{i}$, from equation~\ref{eq:pdf} now refers to the sum of individual sources. Assuming that $M$ is the number of sources to stack, the signal density distribution can be rewritten as follows

\begin{equation}
\label{eq:stacking}
\mathcal{S}_{i} =\frac{ \sum^{M}_{j=1} W_{j} R_{j}(\gamma)S_{i,j}} { \sum^{M}_{j=1} W_{j}R_{j}(\gamma)}
\end{equation}

\noindent where $W_{j}$ is a relative theoretical weight of the $j^{th}$ source in the catalog and $R_{j}(\gamma)$ is the relative detector acceptance for a source with spectral index $\gamma$. As in the all-sky search and the fixed source search, the only two free parameters in the likelihood maximization are the signal contribution, $n_{s}$, and the spectral index, $\gamma$. The weights, $W_{j}$ are correlated with the theoretical expected neutrino luminosity from that given source based on gamma observations, a uniform model will assume all $W_{j} = 1$. The detector acceptance, $R_{j}(\gamma)$, is tabulated using simulation.

\section{Searches and results}
\label{sec:searches}

In the previous two sections we discussed the methodology followed in the point source analysis. With the method described we performed five different searches using the data of the IceCube 40-string configuration:
\begin{itemize}
\item an all-sky search using the standard point source analysis.
\item a dedicated search based on a list of candidate sources to reduce the penalty from the number of trials.
\item a stacking search using the TeV source list from Milagro as a catalog.
\item a stacking search using a catalog of nearby clusters of galaxies.
\item a stacking search using a catalog of stardust galaxies.

\end{itemize}

\subsection{All-sky search and candidate source list results}

\begin{figure}[ht!]  %%% FIGURE 2 %%%
\begin{center}
\epsfysize=6cm \epsfbox{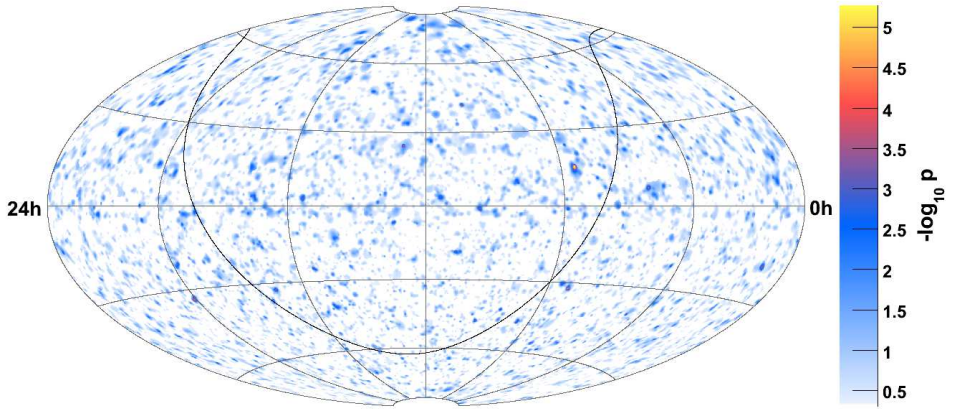} \vspace{0.3cm}
\caption[h]{\label{fig:skymap} Equatorial skymap of {\it pre-trial} significances (p-value) of the all-sky point source. The galactic plane is shown as the solid black curve.}
\end{center}
\end{figure}

The first search is a scan looking for the most significant point in the whole sky. As mentioned before, with the 40-string configuration the search was extended to the southern sky therefore covering declinations from $-85^{\circ}$ to $85^{\circ}$. Figure~\ref{fig:skymap} shows the significance skymap for the all-sky search. The most significant deviation from the only background hypothesis is located at coordinates 113.75$^{\circ}$ r.a., 15.15$^{\circ}$ dec. The {\it pre-trial} estimated p-value of the maximum log likelihood ratio at this location is $5.2 \times 10^{-6}$ equivalent to a single sided sigma of 4.5. 1817 out of 10,000 scrambled data sets have an equal or higher significance somewhere in the sky resulting in the {\it post-trial} p-value of 18\% well compatible with a background fluctuation.

% For Figures insertion uncomment the next section

\begin{table}[htpb]
\begin{center}
\begin{tabular} {c c c c c c c c}
\hline\hline
Object & r.a.~[$^{\circ}$] & dec.~[$^{\circ}$] & $\Phi_{90}$ & p  & n$_s$ & N$_{1^{\circ}}$ & B$_{1^{\circ}}$ \\
\hline
             Cyg OB2 & 308.08 & 41.51 &  5.60 & -- & 0.0 &   2 & 1.8\\
       MGRO J2019+37 & 305.22 & 36.83 &  7.75 & 0.44 & 1.0 &   2 & 1.9\\
       MGRO J1908+06 & 286.98 &  6.27 &  2.83 & -- & 0.0 &   4 & 3.1\\
               Cas A & 350.85 & 58.81 & 10.36 & -- & 0.0 &   1 & 1.8\\
               IC443 &  94.18 & 22.53 &  3.61 & -- & 0.0 &   1 & 2.0\\
             Geminga &  98.48 & 17.77 &  4.15 & 0.48 & 0.7 &   1 & 2.3\\
         Crab Nebula &  83.63 & 22.01 &  3.71 & -- & 0.0 &   1 & 2.0\\
        1ES 1959+650 & 300.00 & 65.15 & 16.05 & -- & 0.0 &   0 & 2.0\\
        1ES 2344+514 & 356.77 & 51.70 &  8.06 & -- & 0.0 &   0 & 1.8\\
               3C66A &  35.67 & 43.04 & 12.76 & 0.24 & 3.5 &   3 & 1.9\\
          H 1426+428 & 217.14 & 42.67 &  6.75 & -- & 0.0 &   3 & 1.8\\
              BL Lac & 330.68 & 42.28 & 11.61 & 0.23 & 2.8 &   3 & 1.8\\
             Mrk 501 & 253.47 & 39.76 &  7.90 & 0.42 & 1.2 &   3 & 2.0\\
             Mrk 421 & 166.11 & 38.21 & 13.00 & 0.14 & 2.6 &   2 & 2.0\\
             W Comae & 185.38 & 28.23 &  4.76 & -- & 0.0 &   0 & 1.9\\
        1ES 0229+200 &  38.20 & 20.29 &  6.62 & 0.18 & 4.1 &   4 & 2.1\\
                 M87 & 187.71 & 12.39 &  3.53 & -- & 0.0 &   2 & 2.5\\
          S5 0716+71 & 110.47 & 71.34 & 18.20 & -- & 0.0 &   0 & 1.6\\
                 M82 & 148.97 & 69.68 & 23.69 & 0.4 & 1.9 &   4 & 1.8\\
            3C 123.0 &  69.27 & 29.67 &  6.53 & 0.44 & 1.3 &   1 & 1.9\\
            3C 454.3 & 343.49 & 16.15 &  3.66 & -- & 0.0 &   1 & 2.3\\
            4C 38.41 & 248.81 & 38.13 &  7.64 & 0.46 & 1.1 &   2 & 2.0\\
        PKS 0235+164 &  39.66 & 16.62 &  6.79 & 0.14 & 5.4 &   5 & 2.3\\
        PKS 0528+134 &  82.73 & 13.53 &  3.83 & -- & 0.2 &   2 & 2.4\\
        PKS 1502+106 & 226.10 & 10.49 &  3.11 & -- & 0.0 &   0 & 2.5\\
              3C 273 & 187.28 &  2.05 &  3.65 & -- & 0.0 &   3 & 3.4\\
            NGC 1275 &  49.95 & 41.51 &  5.66 & -- & 0.0 &   2 & 1.8\\
               Cyg A & 299.87 & 40.73 &  7.95 & 0.44 & 1.2 &   3 & 1.9\\
       IC-22 maximum & 153.38 & 11.38 &  3.15 & -- & 0.0 &   1 & 2.5\\
              Sgr A* & 266.42 & -29.01 & 81.84 & 0.44 & 0.9 &   4 & 3.3\\
        PKS 0537-441 &  84.71 & -44.09 & 123.73 & -- & 0.0 &   3 & 3.5\\
               Cen A & 201.37 & -43.02 & 103.03 & -- & 0.0 &   4 & 3.5\\
        PKS 1454-354 & 224.36 & -35.65 & 93.76 & -- & 0.0 &   4 & 3.5\\
        PKS 2155-304 & 329.72 & -30.23 & 96.64 & 0.32 & 1.3 &   3 & 3.4\\
        PKS 1622-297 & 246.53 & -29.86 & 154.54 & 0.048 & 2.8 &   4 & 3.3\\
        QSO 1730-130 & 263.26 & -13.08 & 24.42 & -- & 0.0 &   4 & 3.5\\
        PKS 1406-076 & 212.24 & -7.87 & 15.52 & 0.39 & 1.6 &   4 & 3.3\\
        QSO 2022-077 & 306.42 & -7.64 & 10.68 & -- & 0.0 &   2 & 3.3\\
               3C279 & 194.05 & -5.79 & 16.14 & 0.12 & 5.9 &   7 & 3.5\\
\hline\hline
\end{tabular}
\caption{\label{tab:sourcelist} Results for the source candidate list. $\Phi_{90}$ is the upper limit of the Feldman-Cousins 90\% confidence interval for an $E^{-2}$ flux, i.e.: $d\Phi/dE \leq \Phi_{90} \, 10^{-12}\mathrm{TeV}^{-1} \mathrm{cm}^{-2} \mathrm{s}^{-1} (E / \mathrm{TeV})^{-2}$.  $n_{s}$ is the best-fit number of signal events; when $n_{s}>0$ the ({\it pre-trial}) p-value is also calculated.  N$_{1^{\circ}}$ is the actual number of events observed in a bin of radius $1^{\circ}$. The background event density at the source declination is indicated by the mean number of background events $B_{1^{\circ}}$ expected in a bin of radius $1^{\circ}$.}
\end{center}
\end{table}

The candidate source list is formed by 39 {\it a priori} selected locations in the sky (see table~\ref{tab:sourcelist}). The most significant results comes from the location of PKS 1622-297 with a final post-trial p-value of 62\% determined as the fraction of scrambled data sets with at least one source with an equal or higher significance. This result is also compatible with the only-background hypothesis.

\subsection{Stacking search results}

For the stacking analysis three different catalogs were chosen. The first search is based on the observations of the Milagro collaboration which reported TeV gamma-ray emission from 16 sources after correlating it with the Fermi Bright Catalog list (Abdo et al. 2009). These sources are promising candidates for neutrino emission (see Beacom \& Kistler 2007 as an example). An additional source was confirmed to have TeV emission, the MGRO J1852+01. We perform a search based on 17 sources observed in TeV gamma rays by Milagro using an equal theoretical weighting from equation~\ref{eq:stacking}.
The second stacked search is performed using 127 starbust galaxies. Starbust galaxies have a dense interstellar medium (ISM) and high star formation rates. The elevated CR rate due to the high number of Super Nova Remnants (SNR) and the dense ISM leads to a potential neutrino emission from these starbust galaxies (see Becker et al. 2009).

Cluster of galaxies are also a potential source of high energy protons and therefore neutrinos. The third stacked analysis uses 5 clusters of galaxies using the model predictions in Murase et al. 2008 as weights in the likelihood.

\begin{table}[hpb]
\begin{center}
\begin{tabular} {c c c c c}
\hline\hline
Catalog &  Model & p-value  & $\Phi_{90} (10^{12}$TeV$^{-1}$cm$^{-2}$s$^{-1})$\\
\hline
  Milagro Sources & $E^{-2}$, Uniform & 0.32 & 9.9 \\
\hline
Starburst Galaxies & $E^{-2}, \propto$ FIR Flux & 1 & 36.2 \\ 
\hline
Clusters of Galaxies  & $E^{-2}$, Uniform & 0.78 &  10.6 \\
\hline\hline
\end{tabular}
\caption{\label{tab:stackingresults} Results for the stacked source searches. The p-value indicates the probability of background to produce the same or higher observed significance. The flux is the upper limit (90\% C.L.) set to the corresponding model.}
\end{center}
\end{table}

The results for the stacking analysis are summarized in table~\ref{tab:stackingresults}. No significant excess has been found and therefore upper limits for a given model have been calculated. In the case of Milagro sources and cluster of Galaxies, the upper limit shown on table~\ref{tab:stackingresults} are calculated assuming a uniform theoretical weighting and an $E^{-2}$ neutrino emission. The upper limits for the starbust galaxies stacking analysis is shown for a model with weighting proportional to the far infrared (FIR) flux at 60~${\mu}m$ since the FIR and radio emission from these sources are correlated with hot dust and a high star formation rate given in Table A.1 in Becker et al. 2009.

\section{Conclusions}
\label{sec:conclusions}

\begin{figure}[ht!]  %%% FIGURE 2 %%%
\begin{center}
\epsfysize=6cm \epsfbox{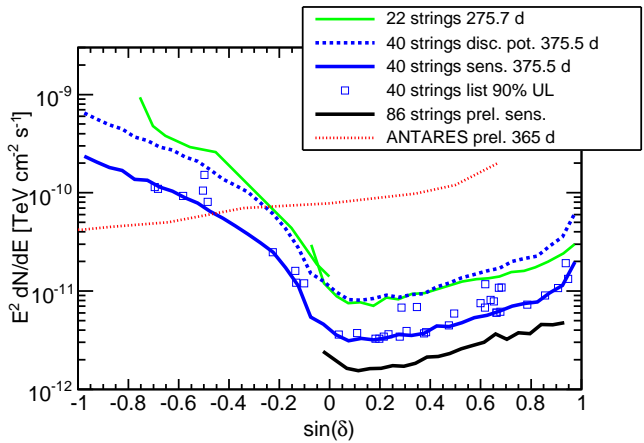} \vspace{0.3cm}
\caption[h]{\label{fig:UL} Sensitivity and upper limits (90\% C.L.) to a point-source $E^{-2}$ $\nu_{\mu}$ flux as a function of declination.
 Blue solid line is the present median sensitivity based on this work. Green line is the previous sensitivity for 22-string configuration of IceCube analysis (Abbasi et al. 2009b, Abbasi et al. 2009c). Future expected sensitivity for IceCube, black solid line, and ANTARES (Coyle, 2010), dotted red line, are also shown. The squares shows the upper limits on the pre-selected 39 direction of the sky for this work.}
\end{center}
\end{figure}

The IceCube observatory has analyzed the data corresponding to the year 2008-9 when the detector configuration consisted of 40 strings. The analysis of time integrated point sources showed no evidence of a neutrino signal and upper limits have been calculated. For an $E^{-2}$ signal of $\nu_{\mu}$ the calculated upper limits from this work are shown in figure~\ref{fig:UL}. The predicted sensitivities of IceCube and the ANTARES neutrino telescope located in the Mediterranean Sea are also shown. The sensitivity is to be interpreted as the median upper limit we expect to observe for a source lying in that particular declination. The current IceCube sensitivity is three times smaller than the previous IceCube 22-string sensitivity (Abbasi et al. 2009b). The upper limits set by IceCube are the best limits on neutrino astronomy so far. A source emitting an equivalent flux to the current sensitivity will be detected by the future IceCube in about 3-5 years, depending on the location in the sky.

\section{Acknowledgment}

We acknowledge the support from the following agencies:
U.S. National Science Foundation-Office of Polar Programs,
U.S. National Science Foundation-Physics Division,
University of Wisconsin Alumni Research Foundation,
U.S. Department of Energy, and National Energy Research Scientific Computing Center,
the Louisiana Optical Network Initiative (LONI) grid computing resources;
National Science and Engineering Research Council of Canada;
Swedish Research Council,
Swedish Polar Research Secretariat,
Swedish National Infrastructure for Computing (SNIC),
and Knut and Alice Wallenberg Foundation, Sweden;
German Ministry for Education and Research (BMBF),
Deutsche Forschungsgemeinschaft (DFG),
Research Department of Plasmas with Complex Interactions (Bochum), Germany;
Fund for Scientific Research (FNRS-FWO),
FWO Odysseus programme,
Flanders Institute to encourage scientific and technological research in industry (IWT),
Belgian Federal Science Policy Office (Belspo);
Marsden Fund, New Zealand;
Japan Society for Promotion of Science (JSPS);
the Swiss National Science Foundation (SNSF), Switzerland;
A.~Gro{\ss} acknowledges support by the EU Marie Curie OIF Program;
J.~P.~Rodrigues acknowledges support by the Capes Foundation, Ministry of Education of Brazil.

%%%%%%%%%%%%%%%%%%%%%%%%%%%%%%%%%%%%%%%%%%%%%%%%%%%%%%%%
% End of the paper
%
\end{document}